%% file: l4dc2026-sample.tex
\documentclass{l4dc2027}

\makeatletter
\def\@jmlrproceedings{}
\def\@jmlrvolume{}
\def\@jmlryear{}
\def\@jmlrpages{}
\makeatother
\title[NL2RCtrl]{NL2RCtrl: Risk-Controlled Controller Synthesis from Natural Language Requirements}
\usepackage{times}
\usepackage{bbm}
\usepackage[most]{tcolorbox}
\usepackage{booktabs}
\usepackage{xcolor}
\usepackage{graphicx}
\usepackage{tikz}
\usetikzlibrary{arrows.meta,positioning,calc,fit,backgrounds}
\coltauthor{%
    \Name{Amir Bayat} \Email{amir.bayat@uclouvain.be}\\
    \Name{Raphael M. Jungers} \Email{raphael.jungers@uclouvain.be}\\
    \addr ICTEAM, UCLouvain, Louvain-la-Neuve, Belgium
}

\begin{document}

\maketitle

\begin{abstract}
Large language models (LLMs) have enabled the
translation of natural language (NL) requirements
into linear temporal logic (LTL) specifications,
facilitating controller synthesis for cyber-physical
systems (CPSs).
However, the reliability of such translations remains
a concern in safety-critical applications, particularly
when NL requirements are ambiguous.
This paper proposes a framework based on Conformal
Risk Control (CRC) to provide a statistical guarantee
on the risk of issuing a controller that violates
the user's intended requirement.
Rather than requiring semantic equivalence between
the generated and intended specifications, the proposed
approach considers language inclusion, a sufficient
condition for ensuring that the traces of a controller
synthesized for the generated specification satisfy
the intended requirement.
To account for ambiguity, the LLM is queried multiple
times, and the resulting LTL specifications are grouped
into semantic equivalence classes and ranked by their
frequency of occurrence. Given a prescribed risk level, CRC selects
how many of the top-ranked equivalence classes
to combine into a single LTL specification,
such that the probability of issuing a controller
that violates the intended requirement
is bounded by that level.
A controller is then synthesized for the resulting
specification, or the framework abstains.
Experiments on a vehicle navigation case study
evaluate the proposed approach.
\end{abstract}

\begin{keywords}
Conformal Risk Control, Formal Controller Synthesis,
Linear Temporal Logic, Large Language Models,
Natural Language Requirements
\end{keywords}

\section{Introduction}

Natural language (NL) is the primary means of human
communication, yet most cyber-physical systems (CPSs)
cannot directly interpret NL requirements.
Translating these requirements into machine-interpretable
representations, such as executable code or temporal
logic (TL), has traditionally required substantial
expert effort. Large language models (LLMs) offer a
promising means of automating this process, but the
inherent ambiguity of NL requirements, together with
the tendency of LLMs to produce plausible yet incorrect
interpretations, raises reliability concerns,
particularly in safety-critical applications
\citep{bubeck2025early}.

\input{Figs/pipeline}

Recent advances in LLM reasoning, including chain-of-thought prompting and problem decomposition \citep{wei2022chain,zhou2023least}, have motivated approaches to synthesizing controllers from NL requirements. Existing methods either generate controller code directly \citep{guo2024controlagent,bayat2025llm} or first translate NL requirements into formal specifications, such as TL, that can be verified and used for controller synthesis \citep{chen2023nl2tl,cosler2023nl2spec,pan2023data,
chen2024autotamp,sundarsingh2025conformalnl2ltl,
bayat2026end,fushimi2026structural,mendoza2026automating}. However, correctness guarantees for synthesized controllers typically assume that the generated specification accurately represents the intended NL requirement, leaving translation errors unaccounted for.

Conformal prediction
\citep{vovk2005algorithmic,angelopoulos2023conformal}
is a statistical framework for constructing prediction
sets that contain the correct response with a prescribed
coverage probability, under the assumption of
exchangeability. Conformal risk control (CRC)
\citep{angelopoulos2024conformal} extends this framework
to user-defined loss functions that are monotone in a
conservativeness parameter, selecting a parameter value
that guarantees the expected loss is bounded by a
prescribed risk level.
These methods have been applied to LLMs,
including conformal prediction sets for generative
outputs \citep{quach2024conformal}, API-based LLMs
\citep{su2024api}, and NL-to-TL translation
\citep{sundarsingh2025conformalnl2ltl}.
More recently, \citet{wang2026scp} employed CRC to
calibrate an accept-or-abstain rule for TL specifications.
Nevertheless, these approaches do not directly address
the risk of issuing a controller synthesized from
an LTL specification that may not accurately reflect
the user's intended requirement.

This motivates the central question of this paper:
Despite the ambiguity inherent in NL requirements,
can controllers be synthesized with statistical
guarantees on the risk of violating the user's
intended specification?
We build on the Abstraction-Based Controller Design
(ABCD) framework of \citet{bayat2026end}, which
provides correctness guarantees relative to a given
TL specification but does not account for uncertainty
in its translation from NL.

To address this limitation, we introduce
NL2RCtrl, a framework illustrated in
Fig.~\ref{fig:framework}.
As in \citet{cosler2023nl2spec}, we query the LLM
multiple times for the same NL requirement and group
the generated TL formulas into semantic equivalence
classes. These classes are ranked according to how
frequently their formulas appear among the LLM outputs.
The highest-ranked semantic equivalence classes are then combined
through conjunction to construct a more conservative
specification. Rather than requiring an exact translation
of the intended TL formula, the proposed approach
relies on semantic implication: a controller synthesized
for the combined specification also satisfies the
intended requirement whenever the former implies the
latter. Increasing the number of combined representative formulas can improve semantic implication, but may also introduce additional constraints that limit controller realizability.

We formulate this tradeoff using CRC to calibrate
the number of top-ranked semantic equivalence
classes to consider, based on a selective loss that
accounts for both semantic implication and controller
issuance. A controller is returned only when
synthesis for the resulting combined specification
succeeds; otherwise, the framework abstains.
Under exchangeability and sound controller synthesis, the resulting procedure guarantees
the following bound for a prescribed
risk level $\alpha$:
\[
\Pr\!\left(
\text{a controller is returned and its closed-loop
behavior violates the intended requirement}
\right)
\leq \alpha.
\]

% This guarantee bounds the probability of issuing
%a controller that violates the intended requirement,
%while allowing the framework to abstain when
%controller synthesis is unsuccessful.

%\textcolor{red}{I would add the contributions in items but I think the two last paragraph make it clear enough.}

%The main contributions of this paper are:
%\begin{itemize}
%    \item We propose a robust NL-to-controller synthesis framework that propagates translation ambiguity to controller synthesis by combining multiple semantically distinct TL interpretations, relaxing the requirement of semantic equivalence to semantic implication.

%    \item We introduce a selective specification loss and apply CRC to calibrate the conservatism of the synthesized specification, establishing a finite-sample statistical guarantee on the risk of incorrect controller issuance.

%    \item We evaluate the framework on task-specific NL-to-TL controller-synthesis benchmarks, investigating the tradeoff between risk and controller issuance, as well as the effects of translation ambiguity and specification conservatism on synthesis performance.
%\end{itemize}

\section{Preliminaries}
\subsection{Linear Temporal Logic and Language Inclusion}

Linear temporal logic (LTL)~\citep{pnueli1977temporal}
is a formal language for specifying temporal properties
of system behavior.
Let $AP$ be a finite set of atomic propositions.
An infinite trace is a sequence
$\sigma=\sigma_0\sigma_1\ldots\in(2^{AP})^\omega$,
where $\sigma_i$ is the set of atomic propositions
that hold at time step $i$.
LTL formulas combine atomic propositions
with temporal operators such as
$\bigcirc$ (\emph{next}),
$\Diamond$ (\emph{eventually}),
$\Box$ (\emph{globally}), and
$\mathcal U$ (\emph{until})
to express how properties are
required to hold over time~\citep[Definition~5.1]{baier2008principles}.

Following the standard language semantics of LTL
\citep[Definition~5.6]{baier2008principles},
the \emph{language} of an LTL formula $\varphi$ is
$\mathcal{L}(\varphi)
=\{\sigma\in(2^{AP})^\omega:\sigma\models\varphi\}$,
i.e., the set of all infinite traces satisfying
$\varphi$.
For two LTL formulas $\varphi_1$ and $\varphi_2$, the relation $
    \mathcal{L}(\varphi_1)
    \subseteq
    \mathcal{L}(\varphi_2)$
is called \emph{language inclusion}. Equivalently, $\varphi_1$ \emph{semantically implies} $\varphi_2$, written $\varphi_1\models\varphi_2$, meaning that every trace satisfying $\varphi_1$ also satisfies $\varphi_2$. Note that language inclusion holds trivially when $\varphi_1$ is \emph{unsatisfiable}, i.e., $\mathcal{L}(\varphi_1)=\varnothing$. If $\mathcal{L}(\varphi_1)\subsetneq\mathcal{L}(\varphi_2)$, we refer to $\varphi_1$ as more \emph{conservative} than $\varphi_2$, since it admits strictly fewer traces. Finally, if $\mathcal{L}(\varphi_1)=\mathcal{L}(\varphi_2)$, the two formulas are \emph{semantically equivalent}, written $\varphi_1\equiv\varphi_2$.

Let $\operatorname{Tr}(\mathcal{S},\pi)$ denote the set
of all traces generated by a system $\mathcal{S}$
under a controller $\pi$.
The resulting closed-loop system satisfies an LTL
formula $\varphi$, denoted by
$(\mathcal{S},\pi)\models\varphi$, if
$\operatorname{Tr}(\mathcal{S},\pi)
\subseteq\mathcal{L}(\varphi)$.
Consequently,
\begin{equation}
    \operatorname{Tr}(\mathcal{S},\pi)
    \subseteq
    \mathcal{L}(\varphi_1)
    \subseteq
    \mathcal{L}(\varphi_2)
    \quad\Longrightarrow\quad
    (\mathcal{S},\pi)\models\varphi_2.
\end{equation}
In other words, if the closed-loop
system satisfies $\varphi_1$, it also satisfies
every formula semantically implied by $\varphi_1$.

%A B\"uchi automaton is a finite-state automaton over infinite
%words, defined as
%$\mathcal{B}=(Q,Q_0,\Sigma,\delta,F)$, where $Q$ is a finite
%set of states, $Q_0\subseteq Q$ is the set of initial states,
%$\Sigma=2^{AP}$ is the input alphabet,
%$\delta\subseteq Q\times\Sigma\times Q$ is the transition relation,
%and $F\subseteq Q$ is the set of accepting states. The size of
%$\mathcal{B}$ is measured as $|Q|+|\delta|$~\cite{baier2008principles}.
%An infinite run is accepting if it visits states in $F$ infinitely often.
%For every LTL formula $\varphi$, a B\"uchi automaton
%$\mathcal{B}(\varphi)$ can be constructed that accepts exactly the traces
%satisfying $\varphi$. We use \texttt{Spot}~\cite{duret2022spot} to
%translate LTL formulas into B\"uchi automata.

\subsection{Abstraction-Based Controller Design}
\label{sec:ABCD}

Consider a CPS with state and input spaces $X$ and $U$, respectively. Abstraction-Based Controller Design (ABCD) constructs a finite symbolic abstraction $\mathcal{S}_a=(X_a,U_a,\rightarrow_a,AP,L)$, where $X_a$ and $U_a$ are finite sets of abstract states and inputs, $\rightarrow_a\subseteq X_a\times U_a\times X_a$ is the transition relation, and $L:X_a\rightarrow 2^{AP}$ is the labeling function. The symbolic abstraction is related to the original CPS through a suitable correctness relation, such as a feedback refinement relation~\citep{reissig2016feedback}. This relation ensures that a controller synthesized for $\mathcal{S}_a$ can be concretized into a controller for the original CPS while preserving satisfaction of the specification.

An LTL formula $\varphi$ can be represented by an equivalent deterministic $\omega$-automaton $\mathcal{A}_\varphi=(Q,2^{AP},\delta,q_0,\operatorname{Acc})$, where $Q$ is a finite set of states, $\delta:Q\times 2^{AP}\rightarrow Q$ is the transition function, $q_0$ is the initial state, and $\operatorname{Acc}$ is the acceptance condition. The form of $\operatorname{Acc}$ depends on the automaton type. A Büchi condition requires a designated set of accepting states to be visited infinitely often. A parity condition assigns an integer priority to each state and accepts a run when the minimum priority occurring infinitely often is even~\citep{baier2008principles,piterman2007nondeterministic, duret2016spot}. In either case, the automaton recognizes exactly the traces that satisfy $\varphi$, i.e., $\mathcal{L}_{\omega}(\mathcal{A}_\varphi)=\mathcal{L}(\varphi)$.

The specification automaton is composed with the
symbolic abstraction through the synchronous product
$S_P=S_a\times A_\varphi$.
Each product state is a pair $(x_a,q)\in X_a\times Q$. The product evolves according to the transition relation of $\mathcal{S}_a$, while the automaton state is updated using the label of the successor abstract state.

%Given an LTL formula $\varphi$ and a control system
%$\mathcal{S}$, the \texttt{Dionysos}
%toolbox~\citep{calbert2024dionysos} constructs a finite
%symbolic abstraction of $\mathcal{S}$, forms its
%synchronous product with the specification automaton,
%and solves a two-player game with a B\"uchi or parity
%acceptance objective.
%The solution yields a \emph{winning set} $W$,
%consisting of the product states from which a control
%strategy can enforce the acceptance condition of the
%specification automaton.
%When $\varphi$ admits an equivalent deterministic
%B\"uchi automaton, the product induces a B\"uchi game;
%otherwise, a deterministic parity automaton is used,
%since deterministic B\"uchi automata cannot represent
%all LTL languages~\citep{baier2008principles}.
%If the initial product state belongs to $W$,
%\texttt{Dionysos} extracts a winning strategy as a
%symbolic controller and concretizes it to the original
%CPS. The resulting closed-loop system then satisfies
%$\varphi$~\citep{tabuada2009verification,calbert2024dionysos}.

The product system induces a two-player game with
a B\"uchi or parity acceptance objective.
Its solution yields a \emph{winning set} $W$,
consisting of the product states from which a
control strategy can enforce the acceptance condition
of the specification automaton.
If the initial product state belongs to $W$,
a winning strategy can be synthesized and
concretized into a controller for the original CPS,
ensuring satisfaction of $\varphi$.

\subsection{Conformal Risk Control}
\label{sec:CRC}

Conformal risk control (CRC) extends conformal prediction
from controlling miscoverage to controlling the expected
value of a user-defined loss function
~\citep{angelopoulos2024conformal}.
Let $Z_1, \ldots,$ $ Z_n$ be random observations
taking values in a space $\mathcal Z$.
These observations form the \emph{calibration set}
$\mathcal D_{\mathrm{cal}}=\{Z_i\}_{i=1}^n$,
while $Z_{n+1}$ denotes a future \emph{test observation}.
CRC requires the calibration and test
observations to be \emph{exchangeable}, meaning that
their joint distribution is invariant under any
permutation of their indices.

Consider a family of decision rules indexed by a
\emph{conservativeness parameter} $k\in\mathcal K$, where
$\mathcal K$ is an ordered set and larger values of $k$
correspond to more conservative decisions.
Let $\ell:\mathcal Z\times\mathcal K\to[0,1]$
be a loss function that is non-increasing in $k$
for every $Z\in\mathcal Z$.
For the CRC guarantee to hold, the most conservative parameter
$k_{\max}\in\mathcal K$ satisfies
$\ell(Z;k_{\max})\leq\alpha$ for every $Z$,
ensuring that the prescribed risk level
$\alpha$ is attainable.

The empirical risk for a given $k$ is
$\widehat{R}_n(k)=\frac{1}{n}\sum_{i=1}^n
\ell(Z_i;k)$.
CRC selects the smallest value of $k$
whose corrected empirical risk does not exceed
a prescribed risk level $\alpha$:
\begin{equation}
    \widehat{k}
    =
    \min\left\{
        k\in\mathcal K:
        \frac{n}{n+1}\widehat{R}_n(k)
        +
        \frac{1}{n+1}
        \leq\alpha
    \right\},
    \label{eq:crc-calibration}
\end{equation}
with the convention that $\widehat{k}=k_{\max}$
if the set is empty.
Under exchangeability and the monotonicity condition,
the calibrated decision rule satisfies
\begin{equation}
    \mathbb{E}\!\left[
        \ell(Z_{n+1};\widehat{k})
    \right]
    \leq\alpha,
    \label{eq:crc-risk}
\end{equation}
where the expectation is taken over
the joint distribution of the calibration and
test observations.

\section{Risk-Controlled NL-to-Controller Synthesis}
\label{sec:method}

Consider a fixed controlled system $\mathcal S$
and a calibration set
$\mathcal D_{\mathrm{cal}}=\{Z_i\}_{i=1}^n$,
where each instance $Z_i=(p_i,\varphi_i^\star)$
consists of an NL requirement $p_i$ and its
corresponding intended LTL specification
$\varphi_i^\star$.
Given a new NL requirement $p$ for the system,
the proposed framework either synthesizes and
returns a controller or abstains, with a statistical bound on the probability of issuing a controller that violates the user's intended specification.

\subsection{Semantic Equivalence Classes and Combined Specifications}
\label{sec:ambiguity-sets}

Given an NL requirement $p$, we query the LLM $m$
times using the prompt provided in
Appendix~\ref{app:prompt}. This produces the candidate
formulas $\varphi^{(1)},\ldots,\varphi^{(m)}$.
We discard any output that is not a syntactically
valid LTL formula, uses atomic propositions outside
the prescribed set $AP$, or is logically unsatisfiable,
i.e., $\mathcal{L}(\varphi^{(j)})=\varnothing$.

The remaining candidate multiset is partitioned into $N(p)$
\emph{semantic equivalence classes}, where two formulas
$\varphi^{(i)}$ and $\varphi^{(j)}$ belong to the same class if
$\varphi^{(i)}\equiv\varphi^{(j)}$.
Each class $\mathcal G$ is assigned the empirical-support score
$s(\mathcal G)=|\mathcal G|/m$, where $|\mathcal G|$ counts the
occurrences of its formulas among the LLM outputs.
Invalid and unsatisfiable outputs remain in the denominator $m$, so the
scores need not sum to one. These scores are used only to rank the
equivalence classes and are not interpreted as calibrated probabilities
of semantic correctness.

We index the equivalence classes by decreasing score, denoting them by
$\mathcal G_1(p),\ldots,\mathcal G_{N(p)}(p)$ with
$s_i(p)=s(\mathcal G_i(p))$ and
$s_1(p)\geq s_2(p)\geq\cdots\geq s_{N(p)}(p)$,
where classes with equal scores are ordered by their first occurrence
among the LLM outputs.
For each $i$, let $\psi_i(p)\in\mathcal G_i(p)$ be an arbitrary LTL
representative. Since all formulas within a class are semantically
equivalent, the choice of representative does not affect the semantic properties
of the class.

Let $K$ be the maximum number of equivalence classes
considered for each NL requirement.
For $k\in\{1,\ldots,K\}$, define
\begin{equation}
    C_k(p)
    =
    \left\{
        \psi_1(p),\ldots,
        \psi_{\min\{k,N(p)\}}(p)
    \right\},
    \qquad N(p)>0.
    \label{eq:ambiguity-set}
\end{equation}
Thus, $C_k(p)$ contains representatives of the $k$
highest-ranked equivalence classes, or all available
representatives if $N(p)<k$.
In particular, $C_k(p)=C_{N(p)}(p)$ for every
$k\geq N(p)$.
If $N(p)=0$, we set $C_k(p)=\varnothing$ and the
framework abstains.

For every nonempty $C_k(p)$, we define the corresponding \emph{combined specification} as
\begin{equation}
    \Phi_k(p)
    =
    \bigwedge_{\psi\in C_k(p)}\psi.
    \label{eq:robust-specification}
\end{equation}
Since $C_k(p)\subseteq C_{k+1}(p)$, it follows that
$\Phi_{k+1}(p)\models\Phi_k(p)$.
Increasing $k$ therefore produces a more conservative
specification by requiring the controller to satisfy
additional candidate formulas.
Although each candidate formula is individually
satisfiable, their conjunction may be unsatisfiable.
In this case, $\mathcal{L}(\Phi_k(p))=\varnothing$,
and no controller can satisfy the resulting specification.

The intended formula $\varphi^\star$ need not appear
among the generated candidates.
For a controller synthesized for $\Phi_k(p)$ to also
satisfy $\varphi^\star$, it is sufficient that
$\Phi_k(p)\models\varphi^\star$.
This condition may hold even when none of the generated
candidates is semantically equivalent to $\varphi^\star$.

\subsection{Controller Synthesis and Selective Loss}
\label{sec:selective-synthesis}

Our objective is to issue controllers that satisfy
the intended specification $\varphi^\star$, while bounding
the probability of issuing one that violates it.
Since $\varphi^\star$ is unknown at deployment, we instead
synthesize controllers for the combined specification
$\Phi_k(p)$.
The parameter $k$ controls the tradeoff between the
conservativeness of the specification and the ability
to synthesize a controller.

\paragraph{Controller synthesis and abstention.}
For each satisfiable $\Phi_k(p)$, we apply ABCD as described in
Section~\ref{sec:ABCD} to compute the winning set
$W_k(p)$. If the prescribed initial condition belongs
to the winning set, we synthesize a controller
$\pi_k(p)$. Otherwise, synthesis fails.
Failure also occurs when no valid candidate is available,
$\Phi_k(p)$ is unsatisfiable, or the synthesis procedure
encounters a computational failure.

Let $\widetilde g_k(p)\in\{0,1\}$ be equal
to 1 if a controller satisfying $\Phi_k(p)$ is
successfully synthesized for $\mathcal S$, and 0 otherwise.
We define the issuance indicator as
\begin{equation}
    g_k(p)
    =
    \prod_{j=1}^{k} \widetilde g_j(p).
    \label{eq:issuance-rule}
\end{equation}
This defines an \emph{absorbing-abstention} policy: once synthesis
fails at any level, the framework abstains at that level and all
subsequent ones. When $g_k=1$, the framework returns
the controller $\pi_k(p)$ and its winning set.

\paragraph{Selective specification loss.}
To quantify the risk of returning an incorrect controller, for a problem instance
$Z=(p,\varphi^\star)$, we define the selective loss
\begin{equation}
    \ell(Z;k)
    =
    g_k(p)
    \mathbbm{1}
    \left[
        \Phi_k(p)\not\models\varphi^\star
    \right].
    \label{eq:selective-loss}
\end{equation}
The loss is one when a controller is issued but
the combined specification does not imply the intended
specification. Abstention incurs zero loss, as no
controller is deployed.

By soundness of ABCD, the synthesized controller satisfies
$\operatorname{Tr}(\mathcal S,\pi_k(p))
\subseteq\mathcal L(\Phi_k(p))$ whenever $g_k=1$.
Consequently, the selective loss upper-bounds the
controller-level error:
\begin{equation}
    g_k(p)
    \mathbbm{1}
    \left[
        \operatorname{Tr}(\mathcal S,\pi_k(p))
        \nsubseteq
        \mathcal L(\varphi^\star)
    \right]
    \leq
    \ell(Z;k).
    \label{eq:controller-loss-bound}
\end{equation}
This bound may be conservative, since failure of semantic
implication does not necessarily imply that the synthesized
controller violates $\varphi^\star$.

\paragraph{Monotonicity of the selective loss.}
Increasing $k$ incorporates additional equivalence classes into
$\Phi_k(p)$, making the specification more conservative.
The following lemma establishes the monotonicity of the
selective loss, which is required for CRC.

\begin{lemma}[Monotonicity of the selective loss]
\label{lem:monotone}
For every problem instance $Z$ and
$k\in\{1,\ldots,K-1\}$, the selective loss satisfies
\begin{equation}
    \ell(Z;k+1)\leq\ell(Z;k).
\end{equation}
\end{lemma}

\begin{proof}
By the definition of the issuance indicator
in~\eqref{eq:issuance-rule},
\[
g_{k+1}(p)
=
g_k(p)\widetilde g_{k+1}(p)
\leq g_k(p).
\]
Moreover, since
$\Phi_{k+1}(p)\models\Phi_k(p)$,
semantic implication is preserved as $k$ increases:
if $\Phi_k(p)\models\varphi^\star$, then
$\Phi_{k+1}(p)\models\varphi^\star$.
Hence,
\[
\mathbbm{1}[
\Phi_{k+1}(p)\not\models\varphi^\star]
\leq
\mathbbm{1}[
\Phi_k(p)\not\models\varphi^\star].
\]
Since both factors in the selective loss are
nonnegative and non-increasing in $k$,
$\ell(Z;k+1)\leq\ell(Z;k)$.
\end{proof}

\subsection{CRC Calibration}
\label{sec:crc-calibration}

We apply CRC to select the number of equivalence classes $k$
such that the expected selective loss is bounded by a prescribed
risk level $\alpha\in(0,1)$. Since the controller issuance rate is non-increasing
in $k$, we seek the smallest value of $k$ satisfying
the risk constraint.

Given a calibration set
$\mathcal D_{\mathrm{cal}}=\{Z_i\}_{i=1}^{n}$, where
$Z_i=(p_i,\varphi_i^\star)$ and the control system
$\mathcal S$ is fixed, we apply the CRC calibration
procedure introduced in Section~\ref{sec:CRC} to select
\begin{equation}
    \widehat{k}
    =
    \min
    \left\{
        k\in\{1,\ldots,K\}:
        \frac{
            \sum_{i=1}^{n}\ell(Z_i;k)+1
        }{n+1}
        \leq\alpha
    \right\}.
    \label{eq:crc-selected-k}
\end{equation}
If the set is empty, we set $\widehat{k}=\infty$,
corresponding to a terminal rule that always abstains,
with $g_\infty\equiv 0$ and $\ell(Z;\infty)=0$.

\begin{theorem}[Risk control]
\label{thm:risk}
Suppose that (i) the calibration and test instances
$Z_1,\ldots,Z_{n+1}$, together with their LLM outputs,
are exchangeable; and (ii) ABCD is sound, i.e.,
every returned controller $\pi_k(p)$ satisfies
$\operatorname{Tr}(\mathcal S, \pi_k(p))
\subseteq\mathcal L(\Phi_k(p))$.
Then, the calibrated parameter $\widehat{k}$ selected
by~\eqref{eq:crc-selected-k} satisfies
\begin{equation}
    \Pr\left(
        g_{\widehat{k}}(p_{n+1})=1
        \ \text{and}\
        \operatorname{Tr}(\mathcal S,
        \pi_{\widehat{k}}(p_{n+1}))
        \nsubseteq
        \mathcal L(\varphi^\star_{n+1})
    \right)
    \leq\alpha.
    \label{eq:risk-guarantee}
\end{equation}
The probability accounts for the randomness in the
calibration and test instances, including the
LLM-generated specifications.
\end{theorem}

\begin{proof}
By Lemma~\ref{lem:monotone}, the selective loss
$\ell(Z;k)$ is non-increasing in
$k\in\mathcal K=\{1,\ldots,K,\infty\}$ and takes
values in $[0,1]$. Moreover, the terminal
abstention rule satisfies $\ell(Z;\infty)=0$.

Thus, \eqref{eq:crc-selected-k}, with
$\widehat{k}=\infty$ when the set is empty,
is the CRC selection rule on $\mathcal K$
with loss bound $B=1$.
Under exchangeability,
\citet[Theorem~1]{angelopoulos2024conformal}
guarantees
$\mathbb E[\ell(Z_{n+1};\widehat{k})]\leq\alpha$.
By soundness of ABCD, the controller-level
error is pointwise bounded by the selective loss,
as established in \eqref{eq:controller-loss-bound}.
Taking expectations establishes
\eqref{eq:risk-guarantee}.
\end{proof}

The guarantee bounds the joint probability of
controller issuance and violation, rather than the
probability of violation conditional on issuance.
Since abstention incurs zero loss, controlling the
risk alone does not ensure frequent controller issuance.
We therefore also evaluate the issuance rate
$\operatorname{Iss}(k)
=\Pr(g_k(p)=1)$.
By the absorbing-abstention rule
in~\eqref{eq:issuance-rule}, the issuance rate is
non-increasing in $k$. Consequently, selecting the
smallest CRC-admissible $k$ maximizes the issuance
rate among the admissible values.

\begin{remark}[Grouped instances]
\label{rem:grouped}
When several NL requirements share the same intended specification, each
instance can instead be defined as
$Z_i=(\{p_{i,r}\}_{r=1}^{M_i},\varphi_i^\star)$, with the selective loss
averaged over its $M_i$ requirements,
$\bar\ell(Z_i;k)=\frac{1}{M_i}\sum_{r=1}^{M_i}
\ell((p_{i,r},\varphi_i^\star);k)$.
Since $\ell$ takes values in $[0,1]$ and is non-increasing in $k$
(Lemma~\ref{lem:monotone}), so does $\bar\ell$. Hence, under
exchangeability of these instances and soundness of ABCD, selecting
$\widehat{k}$ by~\eqref{eq:crc-selected-k} with $\bar\ell$ yields the
joint risk bound~\eqref{eq:risk-guarantee} for $p_{n+1}$ drawn uniformly
from the requirements of $\varphi^\star_{n+1}$.
\end{remark}

\section{Experimental Results}
\label{sec:experiments}

\begin{figure}[t]
    \centering    \includegraphics[width=0.5\linewidth]{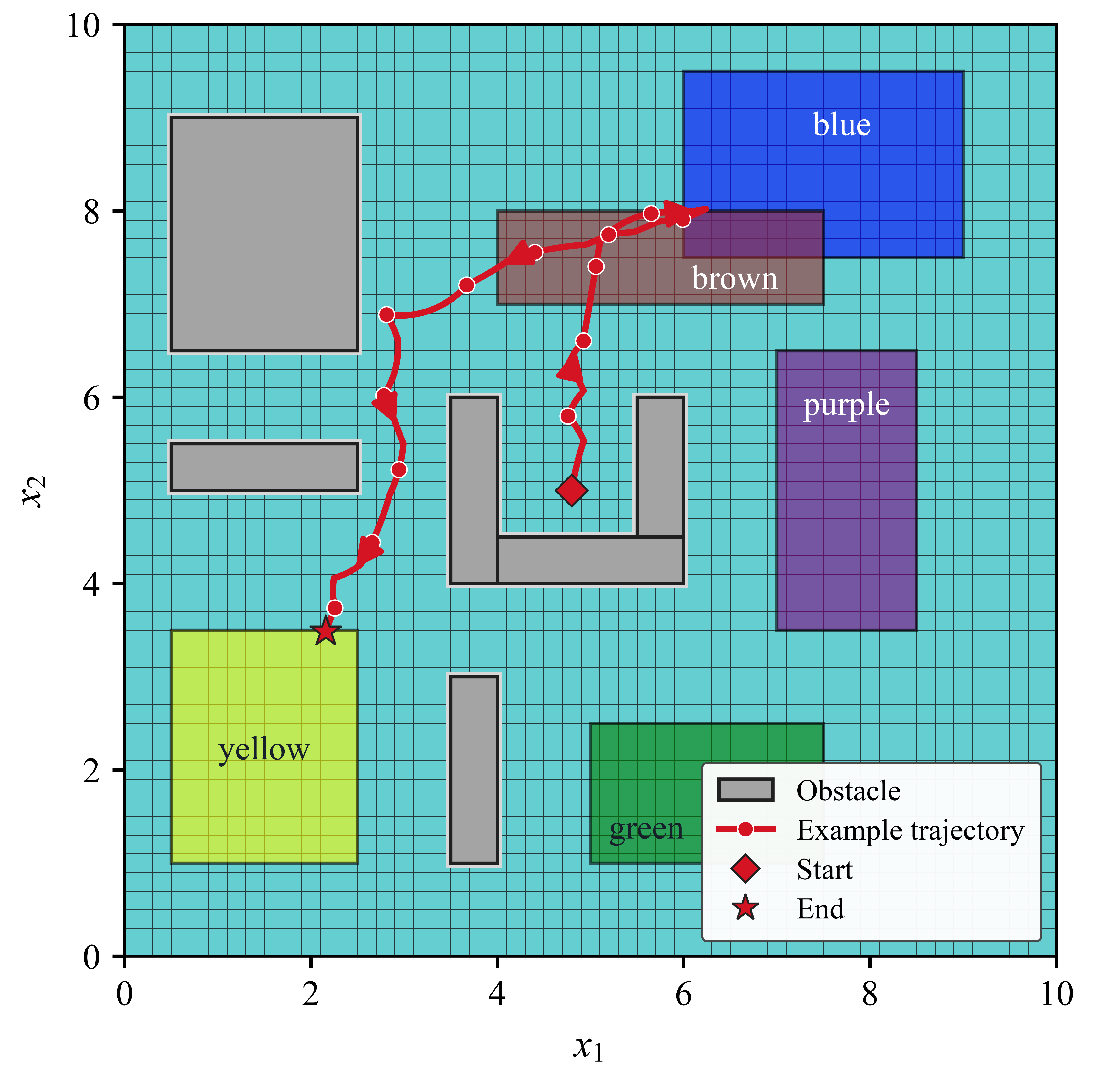}
    \caption{Two-dimensional environment
with colored regions representing atomic propositions,
gray regions representing obstacles, and a grid
illustrating the state-space abstraction.
The red trajectory shows an example of vehicle motion
under a synthesized controller.}
    \label{fig:systme}
\end{figure}

We evaluate the proposed framework using the task-specific
NL-to-LTL benchmark introduced by \citet{bayat2026end},
developed for a vehicle navigation example. The controlled
system, adopted from \citet{reissig2016feedback}, consists of a
vehicle navigating a two-dimensional environment. Following
\citet{bayat2026end}, the environment is partitioned into colored
regions and obstacles, as illustrated in Fig.~\ref{fig:systme}.
The set of atomic propositions is
$AP=\{\texttt{yellow},\allowbreak
\texttt{blue},\allowbreak
\texttt{brown},\allowbreak
\texttt{purple},\allowbreak
\texttt{green},\allowbreak
\texttt{obstacles}\}$,
where each proposition indicates the region occupied by the vehicle.

The benchmark consists of 4{,}736 NL-to-LTL pairs covering 1{,}126
semantically distinct LTL specifications, with a subset of the pairs
manually reviewed for semantic correctness. Using the \texttt{Dionysos}
toolbox~\citep{calbert2024dionysos} for abstraction-based controller
synthesis, controllers were successfully synthesized for 88.10\% of the
reference specifications, while 3.37\% were unrealizable and 8.53\%
exceeded the computational budget.
For evaluation, we randomly partition the 1{,}126 specifications into a
calibration set of 300 and a test set of the remaining 826, assigning all
NL requirements of a specification to the same partition.
Following Remark~\ref{rem:grouped}, each specification and its NL
requirements form one grouped instance, and $\widehat{k}$ is selected
by~\eqref{eq:crc-selected-k} with $\bar\ell$ and $n=300$.
The empirical test risk and issuance rate are computed analogously,
by averaging over the NL requirements of each test specification and
then across the 826 test specifications.

We use DeepSeek V4 Pro to generate $m=10$ candidate LTL specifications per
NL requirement at a sampling temperature\footnote{
The sampling temperature controls the randomness of an LLM's
output distribution, with higher temperatures encouraging
more diverse responses.}
of $\tau=0.7$, selected based on the preliminary evaluation
reported in Section A.2 of the supplementary material.
Since $N(p)\leq m$, we set $K=m=10$, so that every
equivalence class can be included in the combined specification.

\begin{figure}[t]
    \centering
    \begin{minipage}[t]{0.49\linewidth}
        \centering
        \includegraphics[width=\linewidth]{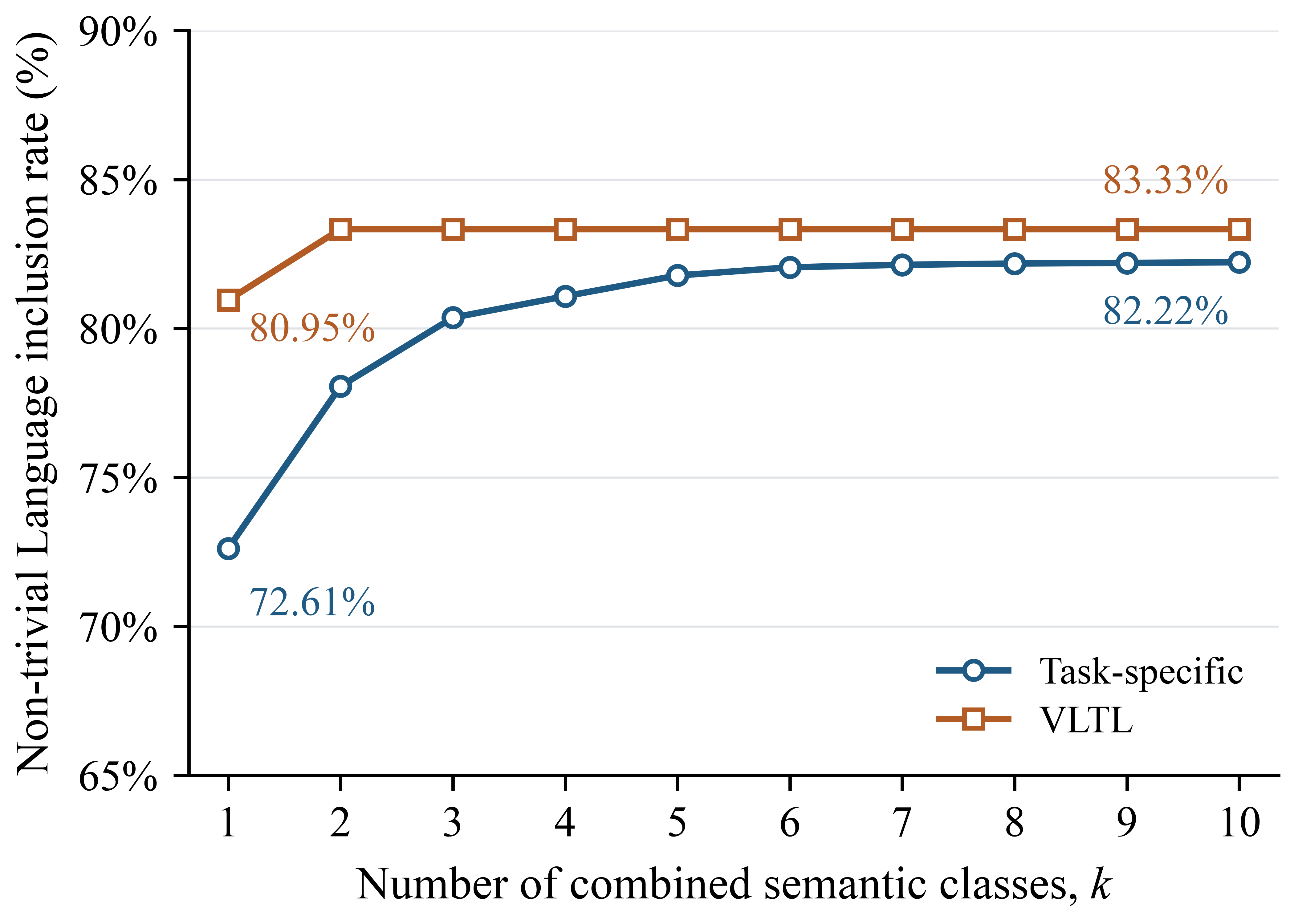}
    \end{minipage}
    \hfill
    \begin{minipage}[t]{0.49\linewidth}
        \centering
        \includegraphics[width=\linewidth]{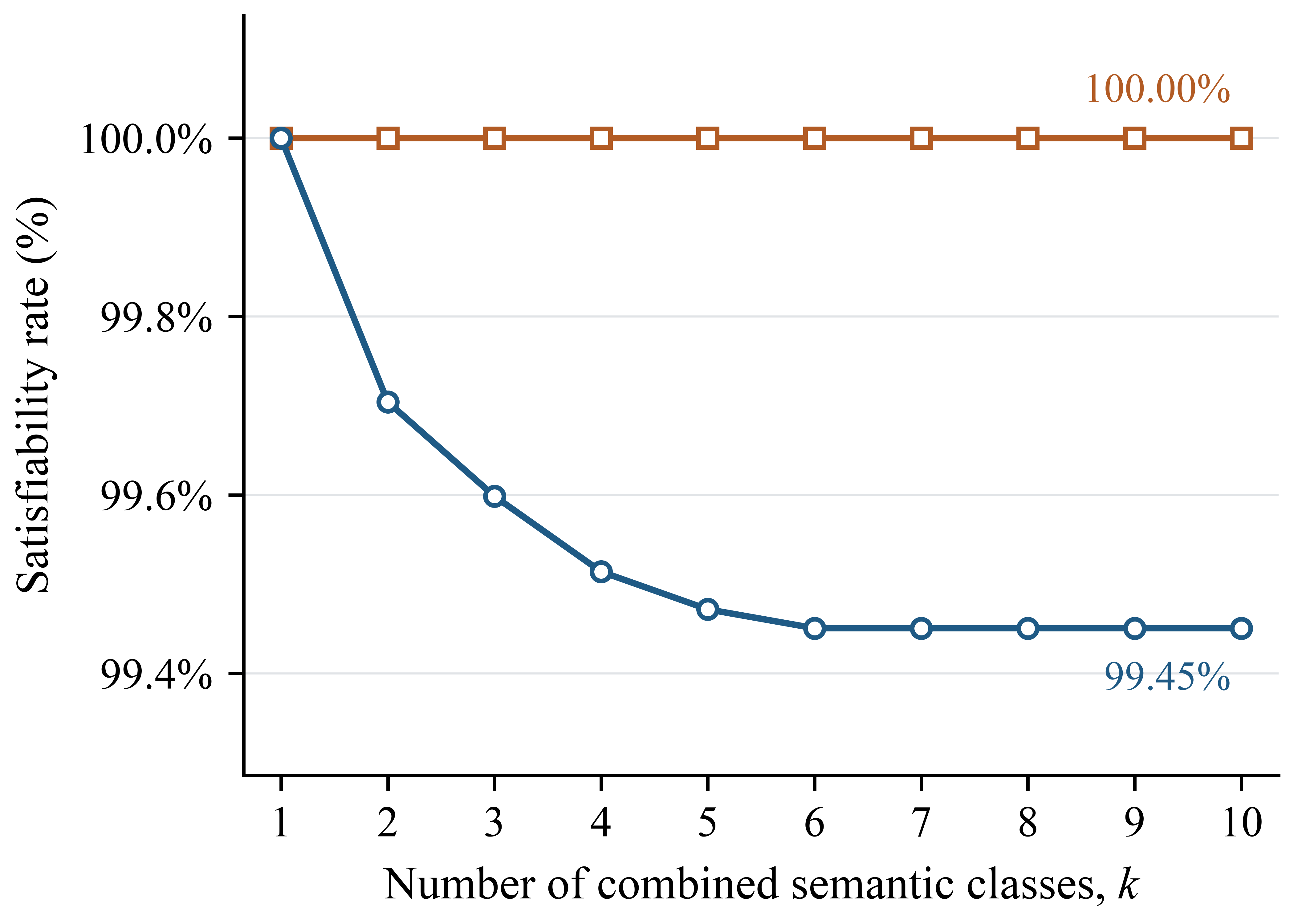}
    \end{minipage}

    \caption{
    Non-trivial language inclusion rate (left) and
    satisfiability rate (right) as functions of the
    number of combined semantic equivalence classes $k$,
    evaluated on the benchmark
    of~\citet{bayat2026end} and
    VLTL-Bench~\citep{english2025verifiable}. The non-trivial language inclusion rate measures the percentage of NL requirements for which the combined specification is satisfiable and implies the intended specification, while the satisfiability rate measures the percentage for which the combined specification remains satisfiable.}
    \label{fig:inclusion_VS_k}
\end{figure}

\paragraph{Effect of Combining Equivalence Classes.}
As shown in Fig.~\ref{fig:inclusion_VS_k} (left), the rate of
non-trivial language inclusion, i.e.,
$\mathcal{L}(\Phi_k(p))\neq\varnothing$ and
$\Phi_k(p)\models\varphi^\star$, increases from
$72.61\%$ to $82.22\%$ as the number of combined
semantic equivalence classes $k$ increases from 1 to 10
on the task-specific benchmark~\citep{bayat2026end}.
On VLTL-Bench~\citep{english2025verifiable},
a general NL-to-LTL benchmark without an associated
controlled system, the inclusion rate increases
more modestly, from 80.95\% to 83.33\%.
The satisfiability rate of $\Phi_k(p)$
(Fig.~\ref{fig:inclusion_VS_k}, right)
remains high, at 99.45\% at $k=10$ on the task-specific benchmark,
while remaining at 100\% on VLTL-Bench.
These results suggest that combining LTL representatives
from multiple semantic equivalence classes can improve
language inclusion while largely preserving
satisfiability across both benchmarks.

\paragraph{Risk-Controlled Controller Synthesis.}
The proposed CRC calibration procedure is evaluated using the selective
loss~\eqref{eq:selective-loss}, averaged over the NL
requirements of each specification (Remark~\ref{rem:grouped}). For each prescribed risk
level $\alpha$, the parameter $\widehat{k}$ is selected
using only the calibration set, and the resulting
controller synthesis outcomes are evaluated on the
test set. Table~\ref{tab:crc_results} reports the selected
$\widehat{k}$, the empirical test risk, and
the controller issuance rate for different prescribed
risk levels $\alpha$.
At $\alpha=0.10$, calibration selects
$\widehat{k}=\infty$, corresponding to abstention for
all requirements. At $\alpha=0.15$, NL2RCtrl selects
$\widehat{k}=3$, achieving an empirical test risk
of 13.49\% and a controller issuance
rate of 84.55\%. As the prescribed risk
level increases, smaller values of $\widehat{k}$
become admissible, resulting in less conservative
combined specifications and higher controller
issuance rates.

Table~\ref{tab:crc_results} also compares NL2RCtrl with a
single LLM query baseline, in which a controller is synthesized directly
from one generated LTL formula, with rates averaged in the same way. The
baseline is evaluated under two correctness criteria: semantic
equivalence, which counts a controller as incorrect if its formula is not
semantically equivalent to the intended specification, and language
inclusion, which counts it as incorrect if its formula does not
semantically imply the intended specification. At an issuance rate of
88.54\%, using language inclusion rather than semantic equivalence as
the correctness criterion reduces the baseline joint risk from 34.52\%
to 22.68\%. Combining equivalence classes with CRC calibration further
reduces the empirical joint risk to 13.49\% at $\alpha=0.15$, at the
cost of a four-percentage-point reduction in issuance (84.55\%).
Unlike the baseline, NL2RCtrl provides a statistical risk guarantee
under the stated assumptions.

%\begin{table}[t]
%\centering
%\caption{CRC calibration results and empirical
%controller-synthesis performance on the test set
%for different risk tolerances $\alpha$.}
%\label{tab:crc_results}
%\small
%\setlength{\tabcolsep}{7pt}
%\begin{tabular}{cccc}
%\toprule
%$\alpha$ & $\widehat{k}$ & Test risk (\%) & Issuance (\%) \\
%\midrule
%%0.05 & $\infty$ & 0 & 0 \\
%0.10 & $\infty$       & 0 & 0 \\
%0.15 & 3       & -- & -- \\
%0.20 & 2       & 16.40 & 86.33 \\
%0.25 & 1        & 22.89 & 89.54 \\
%\bottomrule
%\end{tabular}
%\end{table}

\begin{table}[t]
\centering
\caption{Empirical test risk and controller issuance
for the single LLM query baseline and the proposed
NL2RCtrl framework at different risk levels $\alpha$.}
\label{tab:crc_results}
\small
\setlength{\tabcolsep}{3pt}
\begin{tabular}{lccccc}
\toprule
Method & Statistical guarantee & $\alpha$ & $\widehat{k}$
& Test risk (\%) & Issuance (\%) \\
\midrule
Single LLM query (equivalence) & No & -- & -- &
34.52 & 88.54 \\
Single LLM query (language inclusion) & No & -- & -- &
22.68 & 88.54 \\
\midrule
NL2RCtrl & Yes & 0.10 & $\infty$ & 0 & 0 \\
         & & \textbf{0.15} & \textbf{3} &
         \textbf{13.49} & \textbf{84.55} \\
         & & 0.20 & 2 & 16.40 & 86.33 \\
         & & 0.25 & 1 & 22.89 & 89.54 \\
\bottomrule
\end{tabular}
\end{table}

% Insert CRC calibration and test results here,
% followed by a discussion of risk and issuance.

\section{Conclusion}
This paper introduced a framework that leverages
Conformal Risk Control (CRC) to provide a statistical
guarantee on the risk of issuing a controller that
does not satisfy the user's intended requirement,
even when that requirement is ambiguously expressed.
The proposed method combines multiple semantically
distinct LTL interpretations into a single specification
and calibrates the number of interpretations to be
combined using CRC. Experimental results indicate
that combining multiple interpretations improves
language inclusion, while CRC enables risk-controlled
controller synthesis with the option to abstain.

Nevertheless, combining multiple LTL interpretations
can result in increasingly conservative specifications,
potentially limiting controller realizability.
For stringent risk levels, CRC may select a policy
that always abstains, preventing controller issuance
altogether.
Future work will investigate more
effective strategies for selecting and combining
representative LTL formulas to reduce unnecessary
conservatism while preserving statistical risk
guarantees.

\subsection*{Generative AI Use}

Generative AI tools assisted with language refinement
and experimental code development. The authors verified
all AI-assisted content and take full responsibility
for the manuscript.

\acks{
Amir Bayat is supported by the French Community of Belgium
through an FNRS/FRIA grant.
Raphael M. Jungers is an honorary Research Associate of the FNRS.
This work has received funding from the European Union's
Horizon Europe programme under Grant Agreement No.~101177842
(UniMaaS), the Walloon Region through the FLEXtest project,
and the French Community of Belgium through the ARC project
SIDDARTA.
}

\bibliography{refs}

\appendix

\section{Additional Experimental Details}
\label{app:experiments}

\noindent\textbf{Code Availability.}
The implementation and experimental results are available at
\url{https://github.com/dionysos-dev/Dionysos.jl/tree/NL2RCtrl_L4DC_2027_Amir_B}.

\subsection{NL-to-LTL Prompt}
\label{app:prompt}

The following prompt was used for NL-to-LTL translation.
The natural-language requirement is inserted into the
designated field.

\begin{tcolorbox}[
    colback=gray!3,
    colframe=black!65,
    boxrule=0.5pt,
    arc=2pt,
    left=7pt,
    right=7pt,
    top=7pt,
    bottom=7pt,
    breakable,
    fontupper=\footnotesize
]

You translate one natural-language requirement into a
Linear Temporal Logic (LTL) formula.

\medskip
\textbf{Task atomic propositions:}

\texttt{blue, brown, green, purple, yellow}

Each color proposition is true exactly when the robot
is in that color's region at the current time.
Use only these five color names as atomic propositions
in the output.

\medskip
\textbf{Semantics:}

\begin{itemize}
\setlength{\itemsep}{1pt}
\setlength{\parskip}{0pt}
\setlength{\leftmargin}{1em}
\item Use standard LTL over infinite traces, evaluated
from the current time.
\item \texttt{U} is strong Until:
\texttt{(a U b)} requires \texttt{b} to eventually hold,
while \texttt{a} holds at every preceding time.
\item ``Whenever,'' ``every time,'' and ``globally''
introduce a recurring global requirement.
\item ``From then on'' means that the consequent must
hold globally from the triggering time.
\item Do not add a global operator unless the
requirement expresses a global or recurring condition.
\end{itemize}

\medskip
\textbf{Allowed syntax:}

Atomic propositions:
\texttt{blue, brown, green, purple, yellow}.

Negation: \texttt{!};
Conjunction: \texttt{\&};
Disjunction: \texttt{|};
Implication: \texttt{->};
Equivalence: \texttt{<->};
Next: \texttt{X};
Eventually: \texttt{F};
Globally: \texttt{G};
Until: \texttt{U}.

\medskip
\textbf{Translation rules:}

\begin{itemize}
\setlength{\itemsep}{1pt}
\setlength{\parskip}{0pt}
\item Preserve the complete logical and temporal meaning.
\item Use the exact lowercase color names listed in AP.
\item Do not introduce new propositions.
\item Do not introduce domain assumptions, such as
mutual exclusion, reachability, or ordering, unless
explicitly stated.
\item Do not use every available proposition unless
the requirement needs it.
\item Use parentheses to make operator scope unambiguous.
\item Before answering, silently check temporal scope,
ordering, eventuality, and whether each Until obligation
is strong.
\end{itemize}

\medskip
\textbf{Output rules:}

\begin{itemize}
\setlength{\itemsep}{1pt}
\item Output exactly one LTL formula.
\item Do not output explanations, labels, comments,
or code fences.
\end{itemize}

\medskip
\textbf{Examples:}

\smallskip
\textit{Natural-language requirement:}

Whenever the robot is in the blue region, it must remain
in the yellow region from then on.

\textit{LTL:} \texttt{G((blue) -> G(yellow))}

\smallskip
\textit{Natural-language requirement:}

Whenever the robot is in the blue region, it must stay
in the green region until it reaches the yellow region.

\textit{LTL:} \texttt{G((blue) -> ((green) U (yellow)))}

\smallskip
\textit{Natural-language requirement:}

Eventually reach the blue region, and from that time
eventually reach the yellow region.

\textit{LTL:} \texttt{F((blue) \& F(yellow))}

\smallskip
\textit{Natural-language requirement:}

The robot must never enter the purple region.

\textit{LTL:} \texttt{G(!(purple))}

\smallskip
\textit{Natural-language requirement:}

Stay in the blue region until reaching the brown region.

\textit{LTL:} \texttt{(blue) U (brown)}

\smallskip
\textit{Natural-language requirement:}

At the next step, the robot must be in the green region.

\textit{LTL:} \texttt{X(green)}

\medskip
\textbf{Target input:}

\textit{Available atomic proposition set:}

\texttt{blue, brown, green, purple, yellow}

\smallskip
\textit{Natural-language requirement:}

\texttt{<requirement>}

\textit{[Input natural-language requirement]}

\texttt{</requirement>}

\smallskip
\textit{LTL:}

\end{tcolorbox}

Obstacle avoidance is treated as a system-level
safety constraint rather than a task-specific requirement.
For controller synthesis, every LTL specification is
conjoined with $\Box\neg\texttt{obstacles}$.
The proposition \texttt{obstacles} is therefore excluded
from the benchmark requirements to avoid objectives
that require entering obstacle regions, such as
$\Diamond\texttt{obstacles}$, and is consequently
omitted from the NL-to-LTL prompt.

\subsection{Sampling Temperature Selection}
\label{app:temperature}

To investigate the effect of sampling temperature, we randomly
selected 40 NL-to-LTL pairs from the benchmark and evaluated
$\tau\in\{0.3,0.7,1.0\}$. For each temperature, we queried
the LLM $m=10$ times per NL requirement and measured the
syntactic validity of the generated formulas, their semantic
equivalence to the ground-truth specification, and whether
the conjunction of the generated formulas is satisfiable
and implies the intended specification. We also computed
the mean number of semantic equivalence classes obtained
from the $m$ queries across the 40 requirements.

As shown in Table~\ref{tab:temperature}, both $\tau=0.7$
and $\tau=1.0$ achieve the highest conjunction implication
rate of $80\%$, i.e., the proportion of requirements for which
the conjunction is satisfiable and implies the intended
specification ($\bigwedge_{i=1}^{m}\psi_i\models\varphi^\star$).
However, $\tau=0.7$ achieves a higher syntactic validity
rate while maintaining semantic diversity, with an average
of $2.28$ equivalence classes per requirement.
We therefore select $\tau=0.7$ for the experiments.

\begin{table}[t]
    \centering
    \caption{Effect of sampling temperature on LTL generation
    over 40 NL requirements with 10 queries per requirement.
    Conjunction implication denotes the percentage of
    requirements for which the conjunction is satisfiable
    and implies the intended specification.}
    \label{tab:temperature}
    \small
    \begin{tabular}{ccccc}
        \toprule
        Temperature &
        Syntactically valid &
        Mean classes &
        Semantically equivalent &
        Conjunction implication \\
        \midrule
        $0.3$ & $98.5\%$ & $1.73$ & $47.25\%$ & $70.0\%$ \\
        $0.7$ & $98.5\%$ & $2.28$ & $46.00\%$ & $80.0\%$ \\
        $1.0$ & $96.5\%$ & $2.53$ & $44.75\%$ & $80.0\%$ \\
        \bottomrule
    \end{tabular}
\end{table}

\end{document}

%% file: Figs/pipeline.tex
\begin{figure}[t]
\centering
\begin{tikzpicture}[
    scale=0.85,
    transform shape,
    font=\small,
    >=Latex,
    % General box style
    mainbox/.style={
        line width=0.8pt,
        rounded corners=5pt,
        align=center,
        inner sep=6pt
    },
    subbox/.style={
        draw=violet!55!black,
        line width=0.7pt,
        rounded corners=4pt,
        fill=violet!5,
        align=center,
        inner sep=6pt
    },
    % Input and LLM: white for icon compatibility
    inputbox/.style={
        draw=blue!65!black,
        fill=white
    },
    % LTL specification processing
    specbox/.style={
        draw=cyan!50!black,
        fill=cyan!7
    },
    % Offline risk calibration
    crcbox/.style={
        draw=violet!55!black,
        fill=violet!6
    },
    % Controller synthesis and output
    controlbox/.style={
        draw=green!45!black,
        fill=green!7
    },
    % Arrows
    flow/.style={
        -{Latex[length=2mm]},
        line width=0.85pt,
        draw=black!85
    },
    smalllabel/.style={
        font=\scriptsize,
        fill=white,
        inner sep=2pt
    }
]

% ==============================================
% TOP ROW: NL -> LLM -> LTL
% ==============================================

\node[
    mainbox,
    inputbox,
    text width=3.8cm,
    minimum height=2.85cm
] (nl) at (0,0) {
    \includegraphics[width=1.05cm]{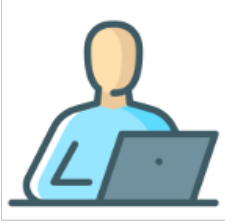}\\[5pt]
    \textbf{NL Requirement}\\[9pt]
    {\scriptsize\color{blue!65!black}
    \textit{$p=$ ``Go to the blue and yellow rooms.''}}\\[7pt]
    {\scriptsize Intended LTL specification}\\[3pt]
    {\scriptsize\color{blue!65!black}
    $\varphi^\star=\Diamond(b\land\Diamond y)$}
};

\node[
    mainbox,
    inputbox,
    text width=2.35cm,
    minimum height=2.85cm
] (llm) at (4.6,0) {
    \includegraphics[width=1.1cm]{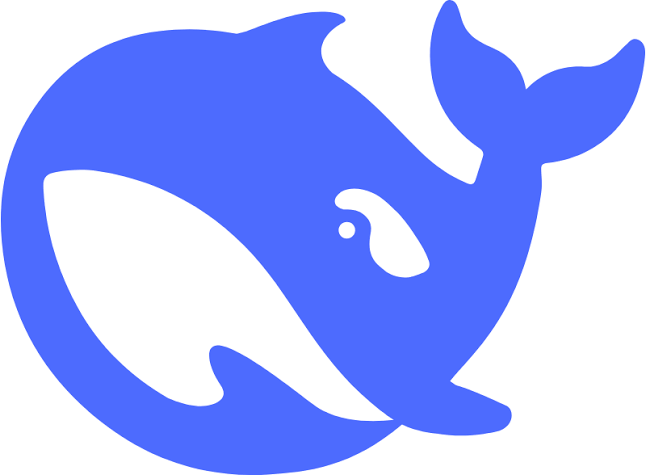}\\[9pt]
    \textbf{LLM}\\[7pt]
    $m$ queries
};

\node[
    mainbox,
    specbox,
    text width=4.0cm,
    minimum height=2.85cm
] (cluster) at (9.15,0) {
    \textbf{Semantic Equivalence Classes and Ranking}\\[6pt]
    {\scriptsize\color{blue!65!black}
    $\psi_1=\Diamond b\land\Diamond y$
    \quad (frequency $s_1$)}\\[5pt]
    {\scriptsize\color{blue!65!black}
    $\psi_2=\Diamond(b\land\Diamond y)$
    \quad (frequency $s_2$)}\\[5pt]
    {\scriptsize\color{blue!65!black}
    $\psi_3=\Diamond(b\land y)$
    \quad (frequency $s_3$)}\\[7pt]
    {\scriptsize Sorted by decreasing frequency:}\\[3pt]
    {\scriptsize $s_1\geq s_2\geq s_3$}
};

\draw[flow] (nl.east) -- (llm.west);
\draw[flow] (llm.east) -- (cluster.west);

% ==============================================
% SECOND ROW: SPECIFICATION -> SYNTHESIS -> OUTPUT
% ==============================================

\node[
    mainbox,
    specbox,
    text width=3.8cm,
    minimum height=1.65cm
] (robust) at (9.15,-3.75) {
    \textbf{Combined Specification}\\[8pt]
    $\displaystyle
    \Phi_{\widehat{k}}(p)
    =
    \bigwedge_{i=1}^{\widehat{k}}
    \psi_i(p)$
};

\node[
    mainbox,
    controlbox,
    text width=3.3cm,
    minimum height=1.65cm
] (abcd) at (4.6,-3.75) {
    \textbf{ABCD Controller Synthesis}\\[6pt]
    System $\mathcal S$ and $\Phi_{\widehat{k}}(p)$
};

\node[
    mainbox,
    controlbox,
    text width=3.8cm,
    minimum height=1.65cm
] (output) at (0,-3.75) {
    \textbf{Synthesis Outcome}\\[7pt]
    Controller $\pi_{\widehat{k}}$ issued\\[2pt]
    or abstention
};

\draw[flow] (cluster.south) -- (robust.north);
\draw[flow] (robust.west) -- (abcd.east);
\draw[flow] (abcd.west) -- (output.east);

% Risk guarantee
\node[
    font=\scriptsize,
    align=center,
    anchor=north,
    text=green!40!black
] at ([yshift=-0.10cm]output.south) {
    $\Pr(\text{incorrect controller issued})\leq\alpha$
};

% ==============================================
% OFFLINE CONFORMAL RISK CONTROL
% ==============================================

\node[
    subbox,
    text width=2.45cm,
    minimum height=0.90cm
] (calset) at (2.90,-6.75) {
    Calibration Set\\[3pt]
    $\mathcal D_{\mathrm{cal}}$
};

\node[
    subbox,
    text width=2.45cm,
    minimum height=0.90cm
] (loss) at (6.70,-6.75) {
    Selective Loss\\[3pt]
    $\ell(Z;k)$
};

\node[
    mainbox,
    crcbox,
    text width=3.15cm,
    minimum height=0.90cm
] (crc) at (4.80,-8.7) {
    \textbf{CRC Calibration}\\[4pt]
    Select $\widehat{k}$
};

% Internal arrows
\draw[flow] (calset.east) -- (loss.west);

\coordinate (lossbend) at
    ($(loss.south)+(0,-0.35cm)$);

\draw[flow]
    (loss.south)
    -- (lossbend)
    -| (crc.north);

% Group background
\begin{scope}[on background layer]
\node[
    draw=violet!55!black,
    line width=0.8pt,
    rounded corners=7pt,
    fill=violet!3,
    fit=(calset)(loss)(crc),
    inner xsep=0.65cm,
    inner ysep=0.50cm
] (crcgroup) {};
\end{scope}

\node[
    font=\small\bfseries,
    text=violet!55!black,
    fill=white,
    inner xsep=7pt
] at (crcgroup.north) {
    Offline Conformal Risk Control
};

% ==============================================
% CALIBRATED PARAMETER -> ROBUST SPECIFICATION
% ==============================================

\draw[flow]
    (crcgroup.east)
    -| node[
        pos=0.55,
        right,
        smalllabel
    ] {$\widehat{k}$}
    (robust.south);

\end{tikzpicture}

\caption{Overview of the NL2RCtrl framework.
Repeated LLM
queries generate candidate LTL formulas, which are
grouped into semantic equivalence classes and ranked
by their frequency of occurrence.
The representatives of the top
$\widehat{k}$ ranked classes form the combined
specification, where $\widehat{k}$ is selected
through offline CRC calibration
to control the risk of issuing
a controller that violates the intended
requirement
while favoring higher controller issuance rates.
At deployment, the framework either synthesizes
and issues a controller or abstains.}

\label{fig:framework}
\end{figure}